\documentclass[aps,prl,preprint,nopacs,superscriptaddress]{revtex4}
\usepackage{amsmath}
\usepackage{amssymb}
\usepackage{graphicx}
\usepackage{hyperref}
\usepackage[utf8]{inputenc}
\newcommand{\beq}{\begin{equation*}}
\newcommand{\eeq}{\end{equation*}}
\newcommand{\topo}{Bi$_2$Se$_3$}
\newcommand{\triv}{In$_x$Bi$_{2-x}$Se$_3$}
\newcommand{\samA}{4QL/4QL 20\%}
\newcommand{\samB}{4QL/2QL 25\%}
\newcommand{\samC}{4QL/2QL 15\%}
\newcommand{\samD}{4QL/1QL 20\%}
\newcommand{\samE}{4QL/1QL 10\%}
\newcommand{\samF}{4QL/1QL 15\%}
\newcommand{\samG}{3QL/3QL 20\%}
\newcommand{\samH}{10QL/10QL 20\%}

\begin{document}

\title{A novel artificial condensed matter lattice and a new platform for one-dimensional topological phases}

\author{Ilya Belopolski} \affiliation{Laboratory for Topological Quantum Matter and Spectroscopy (B7), Department of Physics, Princeton University, Princeton, New Jersey 08544, USA}
\author{Su-Yang Xu} \affiliation{Laboratory for Topological Quantum Matter and Spectroscopy (B7), Department of Physics, Princeton University, Princeton, New Jersey 08544, USA}
\author{Nikesh Koirala} \affiliation{Department of Physics and Astronomy, Rutgers the State University of New Jersey, Piscataway, NJ 08854 USA}
\author{Chang Liu} \affiliation{Department of Physics, South University of Science and Technology of China, Shenzhen, Guangdong 518055, China}
\author{Guang Bian} \affiliation{Laboratory for Topological Quantum Matter and Spectroscopy (B7), Department of Physics, Princeton University, Princeton, New Jersey 08544, USA}
\author{Vladimir N. Strocov} \affiliation{Swiss Light Source, Paul Scherrer Institute, CH-5232 Villigen PSI, Switzerland}
\author{Guoqing Chang} \affiliation{Department of Physics, National University of Singapore, Singapore 117542}
\author{Madhab Neupane} \affiliation{Department of Physics, University of Central Florida, Orlando, Florida 32816, USA}
\author{Nasser Alidoust} \affiliation{Laboratory for Topological Quantum Matter and Spectroscopy (B7), Department of Physics, Princeton University, Princeton, New Jersey 08544, USA}
\author{Daniel Sanchez} \affiliation{Laboratory for Topological Quantum Matter and Spectroscopy (B7), Department of Physics, Princeton University, Princeton, New Jersey 08544, USA}
\author{Hao Zheng} \affiliation{Laboratory for Topological Quantum Matter and Spectroscopy (B7), Department of Physics, Princeton University, Princeton, New Jersey 08544, USA}
\author{Matthew Brahlek} \affiliation{Department of Physics and Astronomy, Rutgers the State University of New Jersey, Piscataway, NJ 08854 USA}
\author{Victor Rogalev} \affiliation{Swiss Light Source, Paul Scherrer Institute, CH-5232 Villigen PSI, Switzerland} \affiliation{Physikalisches Institut and R\"ontgen Center for Complex Material Systems, Universit\"at W\"urzburg, 97074 W\"urzburg, Germany}
\author{Timur Kim} \affiliation{Diamond Light Source, Harwell Campus, Didcot, OX11 0DE, UK}
\author{Nicholas C. Plumb} \affiliation{Swiss Light Source, Paul Scherrer Institute, CH-5232 Villigen PSI, Switzerland}
\author{Chaoyu Chen} \affiliation{Synchrotron SOLEIL, L'Orme des Merisiers, Saint-Aubin-BP 48, 91192 Gif-sur-Yvette, France}
\author{Fran\c{c}ois Bertran} \affiliation{Synchrotron SOLEIL, L'Orme des Merisiers, Saint-Aubin-BP 48, 91192 Gif-sur-Yvette, France}
\author{Patrick Le F\`evre} \affiliation{Synchrotron SOLEIL, L'Orme des Merisiers, Saint-Aubin-BP 48, 91192 Gif-sur-Yvette, France}
\author{Amina Taleb-Ibrahimi} \affiliation{Synchrotron SOLEIL, L'Orme des Merisiers, Saint-Aubin-BP 48, 91192 Gif-sur-Yvette, France}
\author{Maria-Carmen Asensio} \affiliation{Synchrotron SOLEIL, L'Orme des Merisiers, Saint-Aubin-BP 48, 91192 Gif-sur-Yvette, France}
\author{Ming Shi} \affiliation{Swiss Light Source, Paul Scherrer Institute, CH-5232 Villigen PSI, Switzerland}
\author{Hsin Lin} \affiliation{Department of Physics, National University of Singapore, Singapore 117542}
\author{Moritz Hoesch} \affiliation{Diamond Light Source, Harwell Campus, Didcot, OX11 0DE, UK}
\author{Seongshik Oh} \affiliation{Department of Physics and Astronomy, Rutgers the State University of New Jersey, Piscataway, NJ 08854 USA}
\author{M. Zahid Hasan} \affiliation{Laboratory for Topological Quantum Matter and Spectroscopy (B7), Department of Physics, Princeton University, Princeton, New Jersey 08544, USA} \affiliation{Princeton Institute for the Science and Technology of Materials, Princeton University, Princeton, New Jersey 08544, USA} \affiliation{Lawrence Berkeley National Laboratory, Berkeley, CA 94720, USA}


\begin{abstract}
Engineered lattices in condensed matter physics, such as cold atom optical lattices or photonic crystals, can have fundamentally different properties from naturally-occurring electronic crystals. Here, we report a novel type of artificial quantum matter lattice. Our lattice is a multilayer heterostructure built from alternating thin films of topological and trivial insulators. Each interface within the heterostructure hosts a set of topologically-protected interface states, and by making the layers sufficiently thin, we demonstrate for the first time a hybridization of interface states across layers. In this way, our heterostructure forms an emergent atomic chain, where the interfaces act as lattice sites and the interface states act as atomic orbitals, as seen from our measurements by angle-resolved photoemission spectroscopy (ARPES). By changing the composition of the heterostructure, we can directly control hopping between lattice sites. We realize a topological and a trivial phase in our superlattice band structure. We argue that the superlattice may be characterized in a significant way by a one-dimensional topological invariant, closely related to the invariant of the Su-Schrieffer-Heeger model. Our topological insulator heterostructure demonstrates a novel experimental platform where we can engineer band structures by directly controlling how electrons hop between lattice sites.
\end{abstract}

\date{\today}
\maketitle

\section{Introduction}

While crystals found in nature offer a great richness of phenomena, recent progress in physics has also been driven by the study of engineered systems, where key material parameters or Hamiltonian matrix elements can be directly controlled in experiment to give rise to novel emergent properties. For example, the Haldane model for a Chern insulator was recently realized in an optical lattice of ultracold atoms \cite{ColdAtoms} and a Weyl semimetal was observed in a photonic crystal with double-gyroid lattice structure \cite{PhotonicWeyl}. Each advance required engineering a specific ultracold atom optical lattice or photonic crystal. In both cases, these engineered lattices host an emergent band structure, analogous to the band structure formed by electrons in a crystal lattice, but with distinct properties and highly-tuned parameters that gave rise to a novel phase of matter.

Here, we demonstrate an emergent band structure in a novel type of condensed matter lattice based on topological insulators \cite{Mele, HasanKane}. Specifically, we stack together layers of a topological and trivial insulator to create a one-dimensional topological insulator heterostructure. Each interface in the heterostructure hosts a set of topologically-protected interface states which hybridize with each other across the layers, giving rise to a superlattice band structure where the interfaces play the role of lattice sites, the topological surface states play the role of atomic orbitals and the heterostructure acts as a one-dimensional atomic chain. Such a superlattice band structure is emergent in the sense that it arises only when many layers are stacked together, in the same way that an ordinary band structure arises as an emergent property of a crystal lattice. A similar heterostructure has recently been proposed as a simple theoretical model for a Dirac or Weyl semimetal \cite{BurkovBalents}. Here, we experimentally realize and directly observe the first superlattice band structure of this type. We use molecular beam epitaxy (MBE) to build the topological insulator heterostructure and we use angle-resolved photoemission spectroscopy (ARPES) to study its band structure. By adjusting the pattern of layers in the heterostructure, we realize both a topological and a trivial phase, demonstrating that we can tune our system through a topological phase transition. Our work may lead to the realization of novel three-dimensional and one-dimensional symmetry-protected topological phases. At the same time, a topological insulator heterostructure provides a highly-tunable emergent band structure in a true electron system, relevant for transport experiments and device applications.

\begin{figure*}
\centering
\includegraphics[width=16cm]{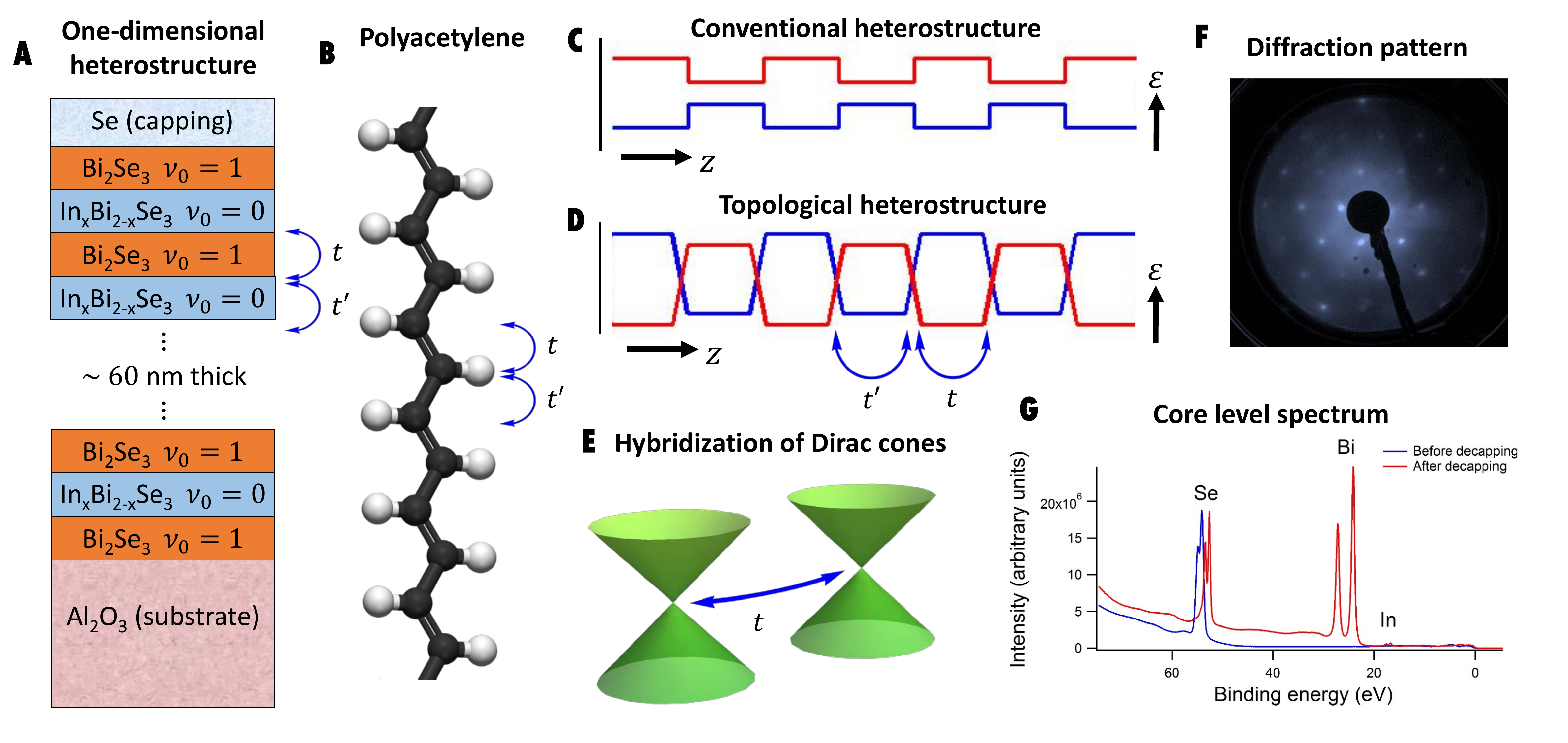}
\caption{\label{Fig1}\textbf{Overview of the topological insulator heterostructure.} (\textbf{A}) The heterostructure consists of a stack of alternating $\mathbb{Z}_2$ topological insulator layers and trivial insulator layers. We use \topo\ as the topologically non-trivial layer and \triv\ as the topologically trivial layer. The $\mathbb{Z}_2$ invariant is denoted by $\nu_0 = 1,0$. An amorphous Se capping layer protects the sample in atmosphere and is removed by heating the sample $\textit{in situ}$. (\textbf{B}) Our system realizes an emergent version of a polyacetylene chain, well-known as a toy model in the study of one-dimensional topological phases. The model has two carbon atoms per unit cell, with one orbital each and hopping amplitudes $t$ and $t'$ associated with the double and single carbon bonds. In the topological insulator heterostructure, the topological and trivial layers play the role of the double and single carbon bonds. (\textbf{C}) A conventional semiconductor heterostructure consists of an alternating pattern of materials with different band gaps. (\textbf{D}) In a topological insulator heterostructure, the band gaps in adjacent layers are inverted, giving rise to topologically-protected Dirac cone interface states between layers. If the layers are thin, then adjacent Dirac cones may hybridize. This hybridization can be described by a hopping amplitude $t$ across the topological layer and $t'$ across the trivial layer. (\textbf{E}) Illustration of the Dirac cone surface states at each interface in the heterostructure, assuming no hybridization. With a hybridization $t$, the topological interface states form a superlattice band structure. (\textbf{F}) A standard low-energy electron diffraction pattern and (\textbf{G}) core level photoemission spectrum show that the samples are of high quality and the Se capping layer was successfully removed by \textit{in situ} heating, exposing a clean sample surface in vacuum.}
\end{figure*}

We first provide a more detailed introduction to the topological insulator heterostructure and argue that our system forms a novel type of condensed matter lattice. A topological insulator heterostructure is built up of alternating layers of topologically trivial and non-trivial insulators. In this work, we use \topo\ as the topological insulator and \triv\ as the trivial insulator, see Fig. \ref{Fig1} A. We note that heterostructures of \topo\ and In$_2$Se$_3$ have already been successfully synthesized \cite{Nikesh, HKU, transport}. Bulk \topo\ has a non-trivial $\mathbb{Z}_2$ invariant, $\nu_0 = 1$ \cite{MatthewBiSe, ZhangBiSe}. Under doping by In, bulk \topo\ undergoes a topological phase transition to a topologically trivial phase with $\mathbb{Z}_2$ invariant, $\nu_0 = 0$ \cite{BrahlekTPT}. In all samples considered here, all \triv\ layers are well into the trivial phase. As a result, the topological invariant flips back and forth from layer to layer in the heterostructure, so that each interface hosts a set of topological interface states \cite{Mele, HasanKane, QiZhang, BAB, HasanMoore, HasanSusu}. Next, we note that these interface states have a finite penetration depth into the bulk of the crystal, so that for sufficiently thin layers, the topological interface states hybridize with each other across the layers and will be subject to an energy level repulsion \cite{MadhabQL, XueQikunQL}. This hybridization can be captured by a hopping amplitude $t$ across the topological layer and $t'$ across the trivial layer. In this way, the topological insulator heterostructure can be viewed as an analog of a polyacetylene chain, shown in Fig. \ref{Fig1} B, where the topological insulator corresponds to the carbon double bond and the trivial insulator corresponds to the carbon single bond (or vice versa). We note that our heterostructure is similar to a conventional semiconductor heterostructure in that the band gap varies in $z$, Fig. \ref{Fig1} C \cite{YuCardona, EsakiTsu}. However, we see that new phenomena arise in a topological insulator heterostructure because the band gap inverts from layer to layer, Fig. \ref{Fig1} D. Specifically, the hybridization of topological interface states, illustrated schematically in Fig. \ref{Fig1} E, gives rise to a superlattice dispersion. In this way, the topological insulator heterostructure is a novel type of condensed matter lattice, where the interfaces correspond to the atomic sites and the topological insulator interface states correspond to the atomic orbitals. Because we can precisely control the thickness of the topological and trivial layers, this superlattice band structure is highly tunable. At the same time, it remains a true electron system, relevant for transport experiments and device applications. Further, if the Fermi level can be placed in the bulk band gap, then the underlying bulk bands of each heterostructure layer become irrelevant, and the transport properties are determined only by the superlattice band structure. This emergent band structure provides a new platform for realizing novel phases in three dimensions, such as a magnetic Weyl semimetal \cite{BurkovBalents}. Further, we note that the energy level repulsion occurs between the electron states at the Dirac points, at $k_{||} = 0$. As a result, the system can be well-described as an atomic chain with a two-site basis and one orbital per site, leading to a two-band single-particle Hamiltonian. By implementing a chiral symmetry in the layer pattern of the heterostructure, it may be possible to realize a phase with strictly one-dimensional topological invariant protected by chiral symmetry, although further theoretical study may be necessary \cite{SSHOriginalPaper, SSHK, PeriodicTable}. We note that prior to measurement by ARPES, we measure standard core level spectra and diffraction patterns to confirm the high quality of our samples, see Fig. \ref{Fig1} F,G.

\section{Results}

We next demonstrate that we have observed an emergent superlattice dispersion in our topological insulator heterostructure. We present a systematic study of four different \topo/\triv\ superlattices, which we write as \samA, \samB, \samC\ and \samD, in Fig. \ref{Fig2} A-D. In this notation, the first parameter refers to the thickness of the \topo\ layer, the second parameter refers to the thickness of the \triv\ layer and the percentage refers to the In doping $x$ of the \triv\ layer. In Fig. \ref{Fig2} E-H, we present ARPES spectra of the four heterostructures along a cut through the $\bar{\Gamma}$ point and in Fig. \ref{Fig2} I-L the same ARPES spectra with additional hand-drawn lines to mark the bands observed in the data. In Fig. \ref{Fig2} M-P, we further present the energy distribution curves (EDCs) of photoemission intensity as a function binding energy at $k_{||} = 0$. These curves correspond to a vertical line on the image plot, indicated by the green arrows in Fig. \ref{Fig2} I-L. The gapless surface state, labeled (1) in \samC\ and \samD, is entirely inconsistent with a \topo\ film 4QL thick. It can only be explained by considering hybridization across the \triv\ layer, demonstrating an emergent superlattice band structure arising from hopping of Dirac cone interface states within the heterostructure. In addition, a gap opens from \samC\ to \samB\ \textit{without any observable bulk band inversion} and \textit {without time-reversal symmetry breaking}. This is an apparent contradiction with the basic theory of $\mathbb{Z}_2$ topological insulators. Again, this result demonstrates a superlattice dispersion. As we increase $t/t'$, the top two lattice sites show larger hybridization and a gap opens in the Dirac cone on the last site. Because ARPES is only sensitive to the top surface of the heterostructure, we cannot observe the inversion of the superlattice bands, but only the increased coupling to the top surface. Our observation of (1) a gapless surface state on a 4QL film of \topo\ and (2) a gap opening in a \topo\ surface state without apparent band inversion each demonstrate that we have observed a superlattice dispersion in our system. In this way, we have shown a completely novel type of electronic band structure, which arises from a lattice of topological insulator Dirac cones which are allowed to hybridize.

We provide a one-dimensional picture of the topological and trivial phases of the superlattice. We consider again \samD, shown in Fig. \ref{Fig3} A-D. Above, we argued that \samD\ has a gapless Dirac cone because $t < t'$. We note that the different hopping amplitude arises from the large bulk band gap and large thickness of the \topo\ layer relative to the \triv\ layer, as illustrated in Fig. \ref{Fig3} E. Alternatively, we can consider how orbitals pair up and gap out in real space. We see that in the topological phase, there are two end modes left without a pairing partner, see Fig. \ref{Fig3} F. We contrast the topological phase with the trivial phase observed in \samA, shown in Fig. \ref{Fig3} G-J. In this case, $t > t'$, because the trivial phase has larger band gap than the topological phase, illustrated in Fig. \ref{Fig3} K. Alternatively, the real space pairing leaves no lattice site without a pairing partner, see Fig. \ref{Fig3} L. We see that our observation of a topological and trivial phase in a topological insulator heterostructure can be understood in terms of an emergent one-dimensional atomic chain where the termination of the chain is either on a strong or weak bond.

We present ARPES spectra of other compositions in the topological and trivial phase to provide a systematic check of our results. First, we compare \samE\ and \samF\ with \samD. The unit cells of these lattices are illustrated in Fig. \ref{Fig4} A-C. The ARPES spectra are shown in Fig. \ref{Fig4} D-F. We also show an EDC at $k_{||} = 0$ in Fig. \ref{Fig4} G-I, as indicated by the green arrows in Fig. \ref{Fig4} D-F. We find that \samE\ and \samF\ also host a gapless surface state and are also topological. This is expected because we have further decreased the In concentration in the trivial layer, increasing the hybridization $t$ and pushing the sample further into the topological phase. Next, we compare \samG\ with \samA. The unit cells are illustrated in Fig. \ref{Fig4} J-L. The ARPES spectra are shown in Fig. \ref{Fig4} M-O and an EDC at $k_{||} = 0$ is shown in Fig. \ref{Fig4} P-R. We find that \samG\ is also in the trivial phase, with a gap in the surface states larger than in \samD. This is consistent with earlier ARPES studies of single thin films of \topo\ on a topologically trivial substrate \cite{MadhabQL, XueQikunQL}. For \samG\ we further observe the same bands at two different photon energies, providing a check that the gapped surface state is not an artifact of low photoemission cross-section at a special photon energy. Our results on \samE, \samF\ and \samG\ provide a systematic check of our results.

\section{Discussion}

Our ARPES spectra show that we have realized the first emergent band structure in a lattice of topological interface states. We have further demonstrated that we can tune this band structure through a topological phase transition. We summarize our results by plotting the compositions of Fig. \ref{Fig2} on a phase diagram as a function of $t'/t$, shown in Fig. \ref{Fig5} A. Our results can also be understood in terms of a one-dimensional atomic chain terminated on either a strong or weak bond. It is natural to ask whether any rigorous analogy exists between our topological insulator heterostructure and a one-dimensional topological phase \cite{PeriodicTable}. An interesting property of our heterostructure is that it naturally gives rise to an approximate chiral symmetry along the stacking direction. In particular, because the wavefunction of a topological insulator surface state decays rapidly into the bulk, we expect that the only relevant hybridization in our heterostructure takes place between adjacent interface states. Further, along the stacking direction, the lattice is bipartite. As a result of nearest-neighbor hopping on a bipartite lattice gives rise to an emergent chiral symmetry along the stacking direction. We note that chiral symmetry is required for a number of topological phases in one dimension, including the BDI, CII, DIII and CI classes, as well as the well-known Su-Schrieffer-Heeger model \cite{PeriodicTable, SSHOriginalPaper, SSHK}. We suggest that future theoretical work could determine whether a one-dimensional topological phase might be realized in our system. Specifically, we might consider topological invariants on the one-dimensional band structure along $k_z$ at fixed $k_{||} = 0$. We also note that we may introduce a second nearest-neighbor hopping by using even thinner layers, see Fig. \ref{Fig5} B. This may break the chiral symmetry, removing the one-dimensional topological invariant and shifting the surface state out of the superlattice bulk band gap. There may further arise an unusual behavior in the surface states if the heterostructure is also characterized by a non-trivial three-dimensional $\mathbb{Z}_2$ invariant. Another application of our topological insulator heterostructure is to break inversion symmetry using a unit cell consisting of four layers of different thickness, see Fig. \ref{Fig5} C. This may lead to a superlattice band structure of spinful bands. More complicated topological insulator heterostructures may allow us to engineer time reversal broken one-dimensional topological phases or large degeneracies implemented by fine-tuning hopping amplitudes. Our systematic ARPES measurements demonstrate the first chain of topological insulator surface states. In this way, we provide not only an entirely novel type of condensed matter lattice but a new platform for engineering band structures in true electron systems by directly controlling how electrons hop between lattice sites.

\section{Materials and Methods}

High quality \topo/\triv\ heterostructures were grown on $10 \times 10 \times 0.5$ mm$^3$ Al$_2$O$_3$ (0001) substrates using a custom-built SVTA-MOS-V-2 MBE system with base pressure of $2 \times 10^{-10}$ Torr. Substrates were cleaned \textit{ex situ} by 5 minute exposure to UV-generated ozone and \textit{in situ} by heating to $800^\circ$C in an oxygen pressure of $1 \times 10^{-6}$ Torr for 10 minutes. Elemental Bi, In, and Se sources, 99.999\% pure, were thermally evaporated using effusion cells equipped with individual shutters. In all samples, the first layer of the superlattice was \topo, 3QL of which was grown at $135^\circ$C and annealed to $265^\circ$C for the rest of the film growth \cite{MBE}. For \triv\ growth, Bi and In were co-evaporated by opening both shutters simultaneously, while the Se shutter was kept open at all times during the growth. To accurately determine the concentration of In in the \triv\ layers, source fluxes were calibrated \textit{in situ} by a quartz crystal microbalance and \textit{ex situ} by Rutherford back scattering, which together provide a measure of the In concentration accurate to within ~1\%. All heterostructures were in total 59QL or 60QL in thickness, depending on the unit cell. In this way, all heterostructures consisted of $\sim 10$ unit cells. All samples were capped at room temperature \textit{in situ} by a $\sim 100$ nm thick protective Se layer to prevent surface contamination in atmosphere. Prior to capping by Se, the high quality of the heterostructure was checked by reflective high-energy electron diffraction (RHEED), see Fig. S1.

Angle-resolved photoemission spectroscopy (ARPES) measurements were carried out at several synchrotron lightsources, in particular at ANTARES, the Synchrotron Soleil, Saint-Aubin, France; ADRESS, the Swiss Light Source (SLS), Villigen, Switzerland; the HRPES endstation of SIS, SLS; I05, Diamond Light Source, Oxfordshire, UK; and CASSIOPEE, Soleil. Samples were clamped onto the sample holder either using an Mo or Ta clamp screwed onto the sample base or thin strips of Ta foil spot-welded onto the sample base. In addition to securely mounting the sample, the clamp also provided electrical grounding. The Se capping layer was removed by heating the sample in a vacuum preparation chamber at temperatures between $200^{\circ}$C and $300^{\circ}$C, at pressures better than $10^{-9}$ Torr, for $\sim 1$hr. Following decapping, the quality of the exposed sample surface was checked by low-energy electron diffraction (LEED) at a typical electron beam energy of 100V. The presence of sharp Bragg peaks and their six-fold rotation symmetry, as shown in main text Fig. 1H, shows the high quality of the exposed sample surface. ARPES measurements were carried out at pressures better than $10^{-10}$ Torr at incident photon energies between 15 eV and 320 eV. We take a short Fermi surface mapping near the center of the Brillouin zone for each sample, at relevant photon energies, to identify the rotation angles corresponding to $\bar{\Gamma}$. This allows us to take $E$-$k$ cuts through the $\bar{\Gamma}$ point to within a rotation angle of $\pm 0.15^{\circ}$, minimizing the error in our measurement of the band gap due to sample misalignment.

\section{Acknowledgments}

Work at Princeton University and synchrotron-based ARPES measurements led by Princeton University were supported by the U.S. Department of Energy under Basic Energy Sciences Grant No. DE-FG-02-05ER46200 (to M.Z.H.). I.B. acknowledges the support of the NSF Graduate Research Fellowship Program. N.K., M.B., and S.O. were supported by the Emergent Phenomena in Quantum Systems Initiative of the Gordon and Betty Moore Foundation under Grant No. GBMF4418 and by the NSF under Grant No. NSF-EFMA-1542798. H.L. acknowledges support from the Singapore National Research Foundation under award No. NRF-NRFF2013-03. M.N. was supported by start-up funds from the University of Central Florida. We acknowledge Diamond Light Source, Didcot, U.K., for time on beamline I05 under proposal SI11742-1. We acknowledge measurements carried out at the ADRESS beamline \cite{ADRESS} of the Swiss Light Source, Paul Scherrer Institute, Switzerland. We acknowledge J. D. Denlinger, S. K. Mo, and A. V. Fedorov for support at the Advanced Light Source, Lawrence Berkeley National Laboratory, Berkeley, CA, USA. C.L. was supported by Grant No. 11504159 of the National Natural Science Foundation of China (NSFC), Grant No. 2016A030313650 of NSFC Guangdong, and Project No. JCY20150630145302240 of the Shenzhen Science and Technology Innovations Committee.

\vspace{0.5cm}

All data needed to evaluate the conclusions in the article are present in the article and/or the Supplementary Materials. Additional data is available from the authors upon request.

\section{Author Contributions}

This work was conceived by I. B., S.-Y. X. and M. Z. H. Samples were grown and characterized by N. K., M. B. and S. O. N. K. and V. S. provided crucial creative insight into sample synthesis and measurement. Synchrotron ARPES measurements were carried out by I. B. with assistance from S.-Y. X., C. L., G. B., M. N., N. A., D. S. and H. Z. Synchrotron ARPES endstations were built and supported by V. S., V. R., T. K., N. P., C. C., F. B., P. F., A. T.-I., M.-C. A., M. S. and M. H. G. C. and H. L. carried out numerical calculations. M. Z. H. provided overall direction, planning and guidance for the project.

\section{Competing Interests}

The authors declare that they have no competing interests.

\clearpage

\clearpage
\begin{figure}
\centering
\includegraphics[width=13cm]{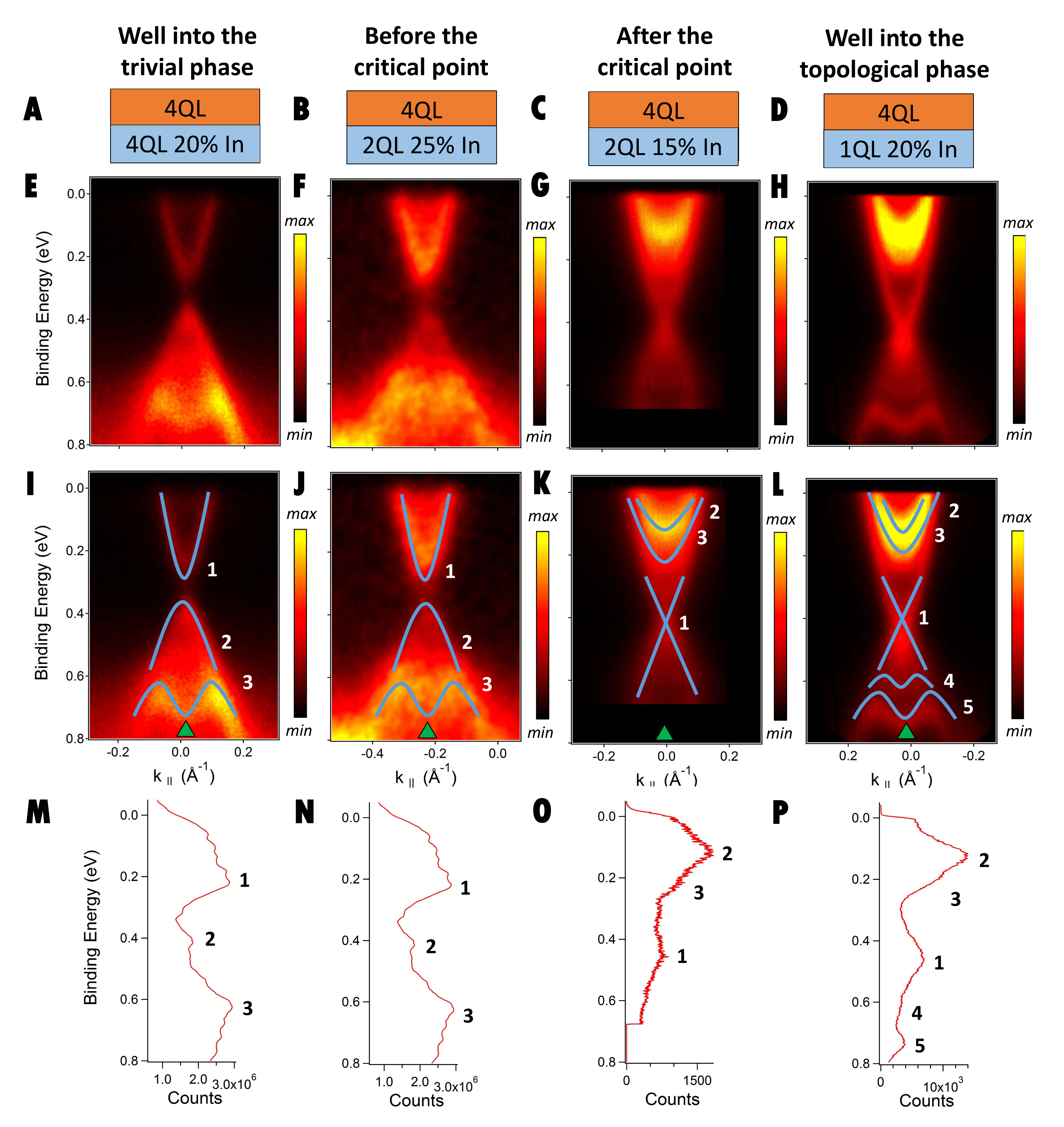}
\end{figure}

\clearpage
\begin{figure}
\caption{\label{Fig2}\textbf{Observation of an emergent superlattice band structure in trivial and topological phases.} (\textbf{A-D}) The unit cells of the heterostructures studied, with different thickness and In-doping of the \triv\ layer. (\textbf{E-H}) ARPES spectra of heterostructures. We see that (\textbf{E}) and (\textbf{F}) showed gapped surface states while (\textbf{G}) and (\textbf{H}) show gapless surface states. Note that in all samples measured, the top layer of the superlattice, which is the only layer directly measured by ARPES, is 4QL of \topo. Nonetheless, the spectra differ strikingly. (\textbf{I-L}) The same spectra as in (\textbf{E-H}), but with additional hand-drawn lines showing the key features of the spectra. In the \samA\ and \samB\ samples, we observe a gapped topological surface state, (1) and (2), along with a valence band quanutm well state (3). By contrast, in \samC\ and \samD, we observe a gapless Dirac cone surface state, (1), two conduction band quantum well states, (2) and (3) and, in \samD, two valence quantum well states, (4) and (5). We emphasize that the gapless Dirac cone is observed even though the top \topo\ layer is only 4QL thick. (\textbf{M-P}) Energy distribution curves through the $\bar{\Gamma}$ point of each spectrum in (\textbf{E-H}). The peaks corresponding to the bulk quantum well states and surface states are numbered based on the correspondance with the full spectra, (\textbf{E-H}).}
\end{figure}

\clearpage
\begin{figure}
\centering
\includegraphics[width=16cm]{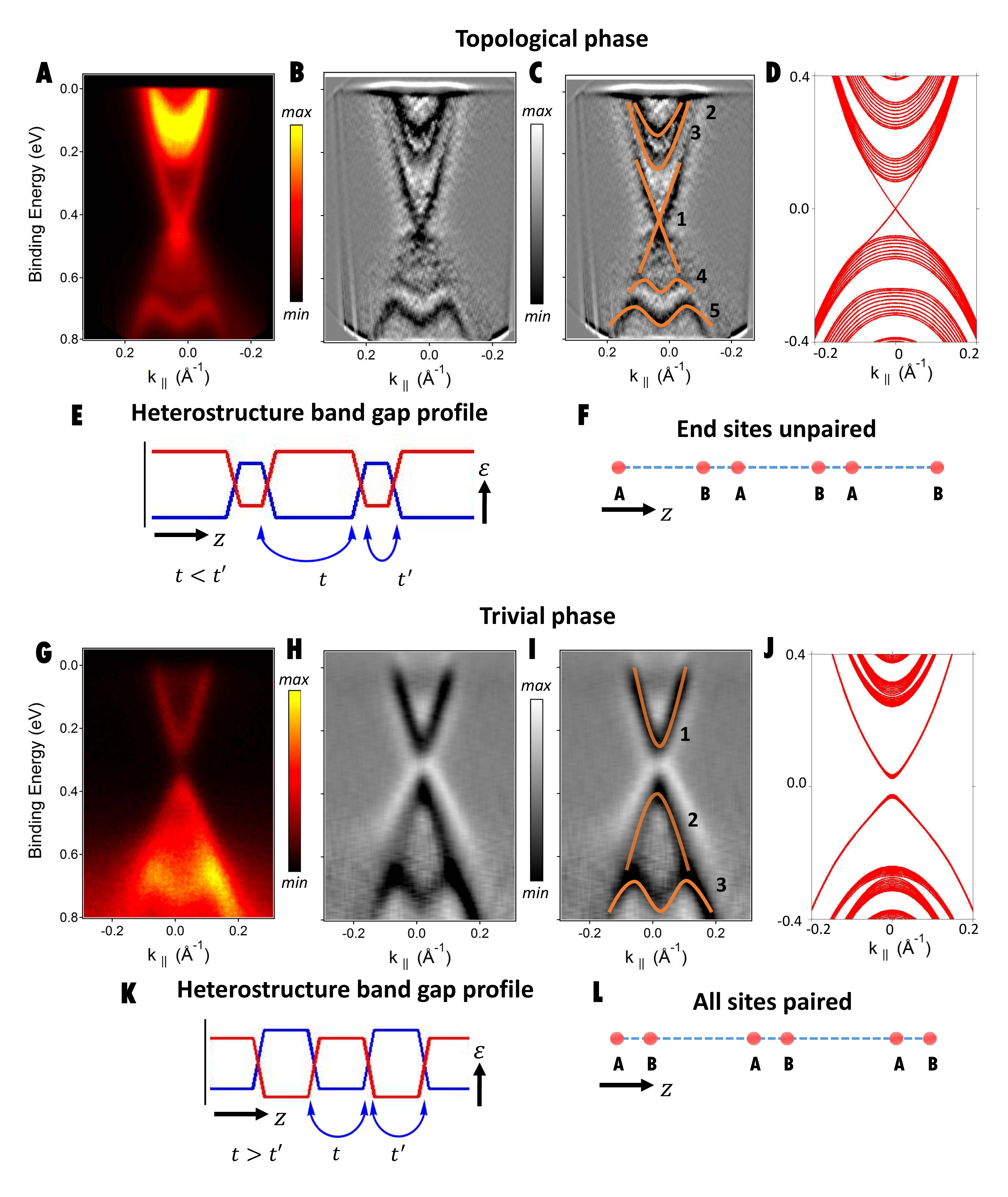}
\end{figure}

\clearpage
\begin{figure}
\caption{\label{Fig3}\textbf{Realization of the Su-Schrieffer-Heeger model.} (\textbf{A}) ARPES spectrum of \samD, with (\textbf{B}) a second-derivative map of the same spectrum and (\textbf{C}) the same second-derivative map with additional hand-drawn lines highlighting the gapless surface state and the two quantum well states in the conduction and valence bands. (\textbf{D}) A tight-binding calculation demonstrating a topological phase qualitatively consistent with our experimental result. (\textbf{E}) A cartoon of the band gap profile of the \samD\ heterostructure. The trivial layer is much thinner than the topological layer, so $t < t'$ and we are in the SSH topological phase. (\textbf{F}) The SSH topological phase can be understood as a phase where the surface states pair up with their nearest neighbors and gap out but where the pairing takes place in such a way that there is an unpaired lattice site at the end of the atomic chain. (\textbf{G}) ARPES spectrum of \samA, with (\textbf{H}) a second-derivative map of the same spectrum and (\textbf{I}) the same second-derivative map with additional hand-drawn lines highlighting the gapped surface state and the quantum well state in the valence band. (\textbf{J}) A tight-binding calculation demonstrating a trivial phase qualitatively consistent with our experimental result. (\textbf{K}) A cartoon of the band gap profile of the \samA\ heterostructure. The trivial layer is as thick as the topological layer, with a larger band gap due to high In-doping, so $t > t'$ and we are in the SSH trivial phase. (\textbf{L}) The SSH trivial phase can be understood as a phase where all lattice sites in the atomic chain have a pairing partner.}
\end{figure}

\clearpage
\begin{figure}
\centering
\includegraphics[width=11cm]{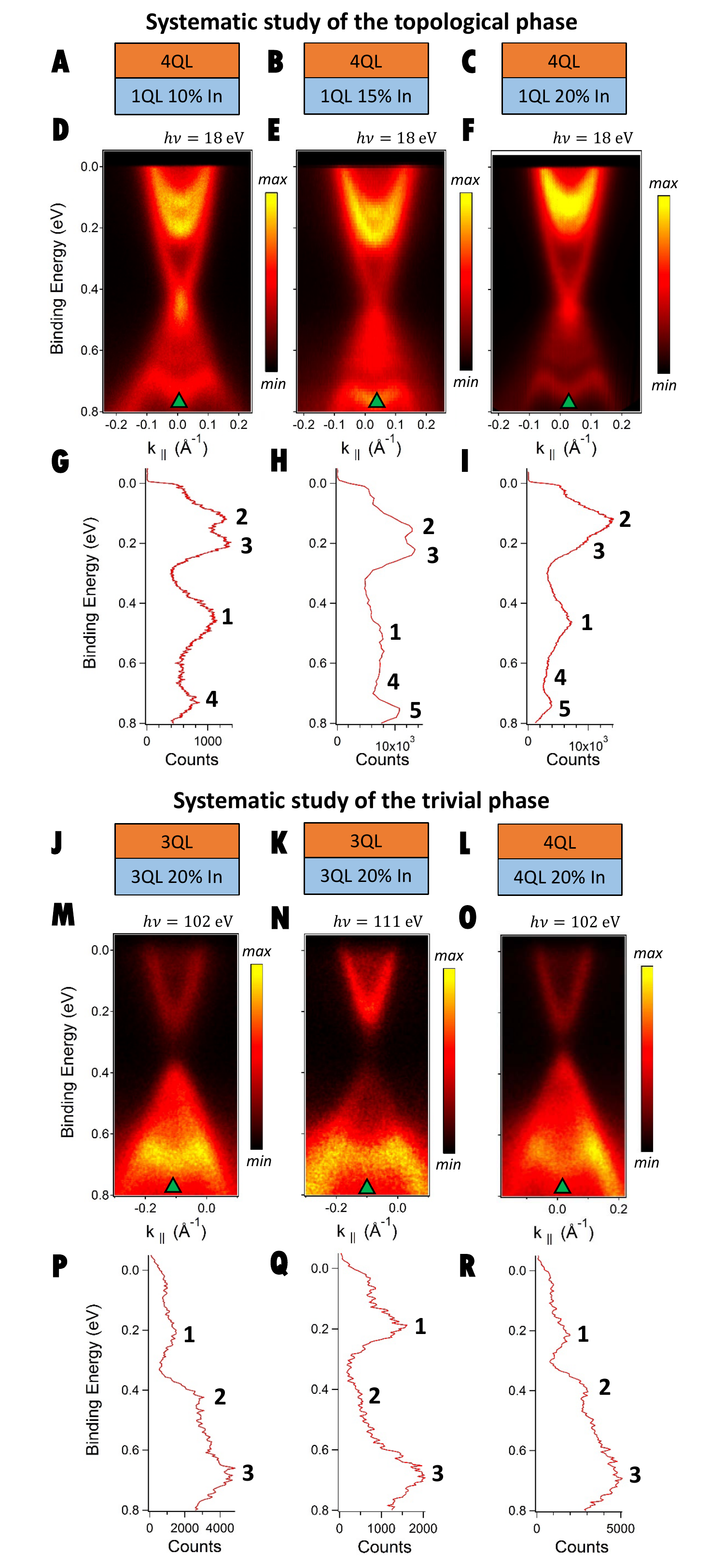}
\end{figure}

\clearpage
\begin{figure}
\caption{\label{Fig4}\textbf{Systematic study of the topological and trivial phases.} (\textbf{A-F}) ARPES spectra of 4QL/1QL samples at $h \nu = 18$ eV with varying concentrations of In, showing a robust topological phase. (\textbf{G-I}) Energy distribution curves through the $\bar{\Gamma}$ point of each spectrum in (\textbf{D-F}). (\textbf{J-O}) ARPES spectra of 3QL/3QL 20\% and 4QL/4QL 20\% samples at $h \nu = 102$ eV and a 3QL/3QL 20\% sample at $h \nu = 111$ eV, showing a robust trivial phase. (\textbf{P-R}) Energy distribution curves through the $\bar{\Gamma}$ point of each spectrum in (\textbf{M-O}).}
\end{figure}

\clearpage
\begin{figure}[h!]
\centering
\includegraphics[width=10cm]{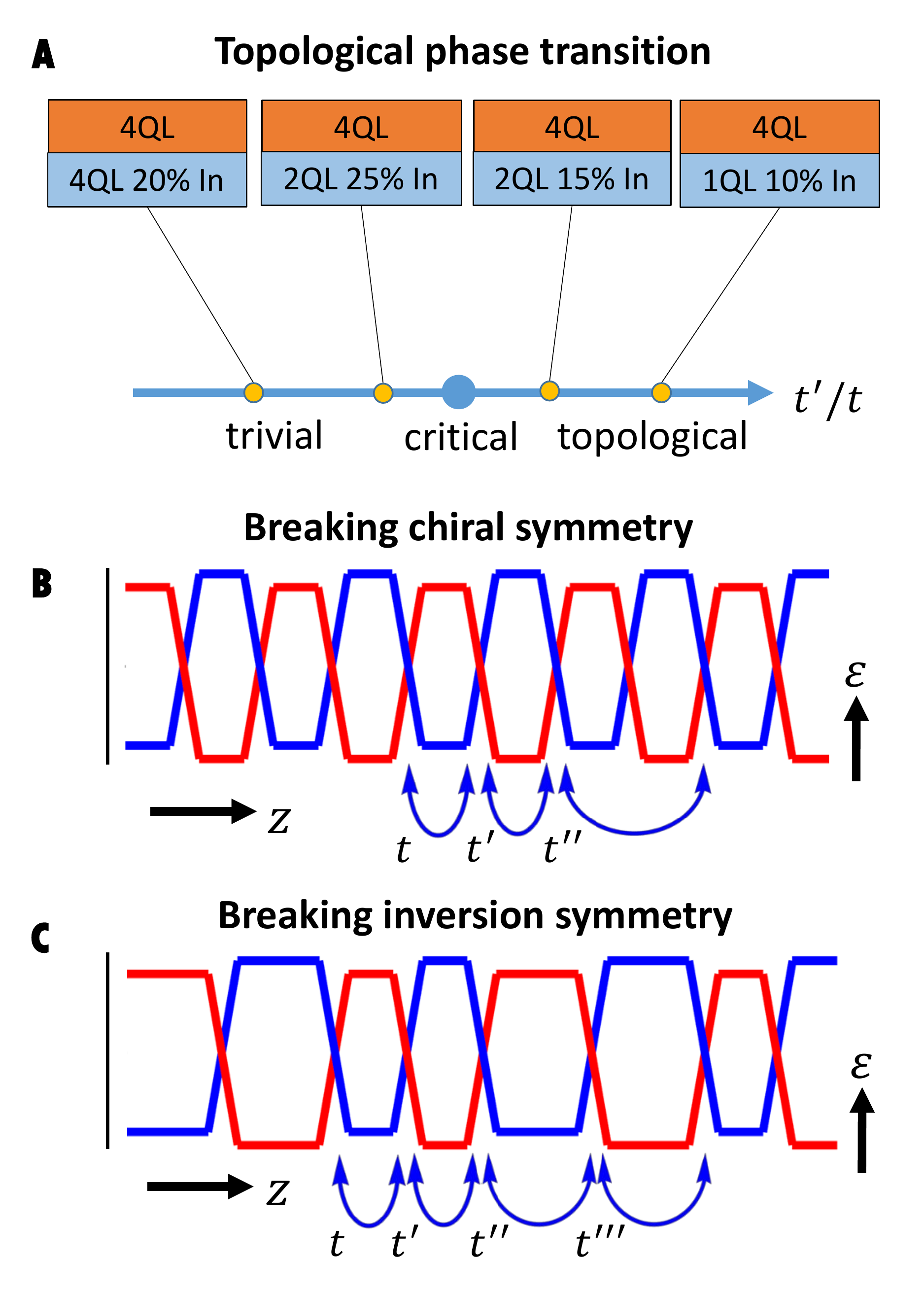}
\caption{\label{Fig5}\textbf{Phase diagram of a tunable emergent superlattice band structure.} (\textbf{A}) By varying the hopping $t'$ across the trivial layer we realize a trivial phase and topological phase in our superlattice band structure. (\textbf{B}) In the heterostructures demonstrated here, we expect that only nearest-neighbor hopping is relevant, giving rise to an emergent chiral symmetry along the stacking direction. By using thinner layers we propose to introduce a next nearest-neighbor hopping, breaking this chiral symmetry. This may change the topological classification of our system. (\textbf{C}) By doubling the unit cell, we can break inversion symmetry. This will give rise to a spinful superlattice dispersion.}
\end{figure}

\clearpage

\begin{center}
\textbf{\large Supplementary Materials}
\end{center}

\setcounter{equation}{0}
\setcounter{figure}{0}
\setcounter{table}{0}
\makeatletter
\renewcommand{\theequation}{S\arabic{equation}}
\renewcommand{\thefigure}{S\arabic{figure}}
\renewcommand{\thetable}{S\arabic{table}}

\section{1. Estimate of Indium diffusion in the heterostructure}

\begin{figure}[h!]
\centering
\includegraphics[width=15cm]{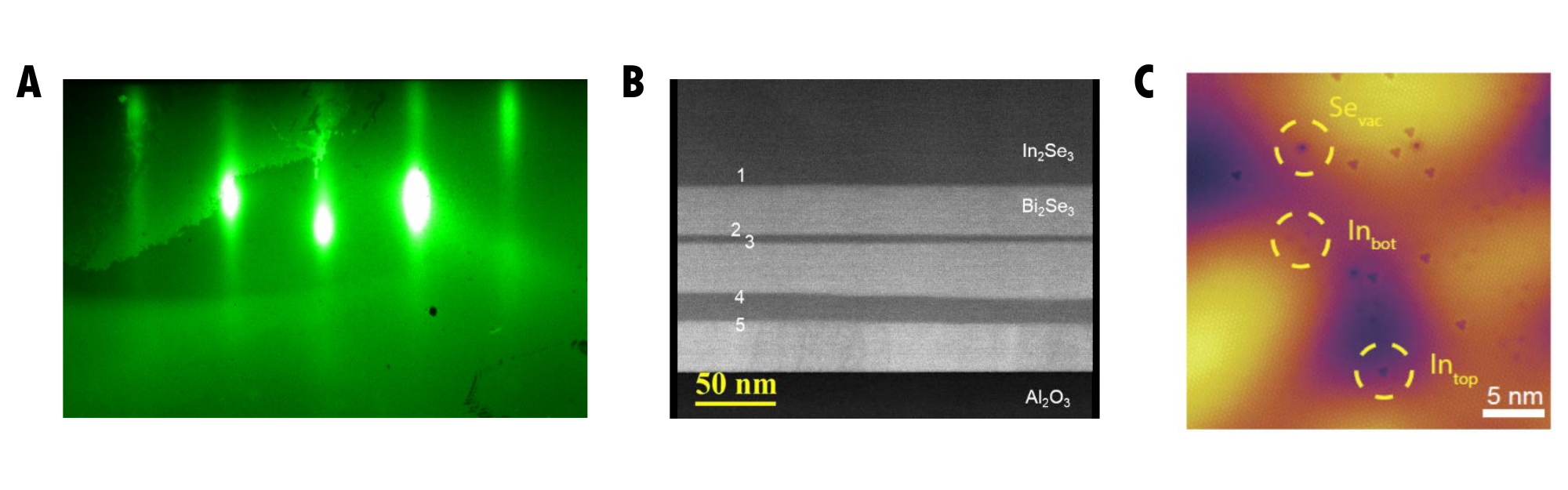}
\caption{\label{FigS4}\textbf{Characterization of In diffusion in the heterostructures.} (\textbf{A}) Reflection high-energy electron diffraction (RHEED) image showing an atomically flat single crystalline growth, as indicated by the sharp streaks and Kikuchi lines. This confirms the high quality of our heterostructures. (\textbf{B}) High-angle annular dark-field scanning transmission electron microscopy (HAADF-STEM) image of a \topo/In$_2$Se$_3$ heterostructure, showing sharp interfaces between the layers. The sample has layer pattern 100 In$_2$Se$_3$ / 30 \topo\ / 5 In$_2$Se$_3$30 / 30 \topo\ / 20 In$_2$Se$_3$ / 30 \topo, demonstrating at the same time the sharp interfaces in our heterostructure as well as our ability to precisely control the layer thickness in MBE. (\textbf{C}) Indium atoms observed on \topo\ by scanning tunneling microscopy (STM) in a 5QL/20QL 50\% heterostructure with one unit cell. In$_\textrm{top}$ and In$_\textrm{bot}$ represent In atoms diffused to the top and bottom Bi layer within the topmost QL of \topo, while Se$_\textrm{vac}$ represent Se vacancies. (Reprinted from Ref. [24], $\copyright$ 2015 American Chemical Society).}
\end{figure}

It is important to check that diffusion of In within the sample does not remove the layered \topo/\triv\ pattern of the heterostructure. In this section, we characterize In diffusion in our heterostructures. In subsequent sections, we also provide a number of checks in ARPES which confirm indirectly that the heterostructure consists of clean interfaces. We show a high-angle annular dark-field scanning transmission electron microscopy (HAADF-STEM) image, Fig. \ref{FigS4} B, of a thin film with alternating layers of \topo\ and In$_2$Se$_3$ of different thicknesses [9]. Sharp interfaces between the In$_2$Se$_3$ and Bi$_2$Se$_3$ indicate no substantial diffusion between the layers. We also measure a heterostructure consisting of a single \topo/\triv\ unit cell, with the \topo\ on top, see Fig. \ref{FigS4} C. We use scanning tunneling microscopy (STM) to count the number of In atoms that diffuse to the top QL of Bi$_2$Se$_3$, see Ref. 24 for details of the STM measurement. For both 30QL/20QL 50\% and 5QL/20QL 50\% films grown at $275^{\circ}$C, we find $\sim 0.2 \%$ In diffusion to the top QL of \topo\ [24]. This indicates that there is minimal In diffusion at $x = 50\%$ and also that there is no significant variation of In diffusion with the thickness of the \topo\ layer. To gain insight into the effect of growth temperature and doping $x$, we further studied 3QL/20QL 50\% grown at $300^{\circ}$C and 30QL/20QL 100\% grown at $275^{\circ}$C. We found $\sim 2\%$ and $\sim 1.3\%$ In diffusion to the top QL of the \topo\ layer, respectively (data not shown). Clearly (and perhaps not unexpectedly) In diffusion is suppressed with lower growth temperature and lower $x$. Since the heterostructures studied here are grown at $265^{\circ}$C with $x \leq 25\%$, In diffusion should be considerably suppressed even compared to the $\sim 0.2 \%$ diffusion observed for 30QL/20QL 50\% and 5QL/20QL 50\% grown at $275^{\circ}$C. Finally, we note that the topological phase transition for \triv\ occurs at $x \sim 4\%$, so we can conclude that In diffusion in our heterostructures is easily low enough to ensure that the \topo\ and \triv\ layers remain topological and trivial, respectively. Our direct characterization of In diffusion in our samples shows that In diffusion does not remove the layered pattern of the heterostructure.

\section{2. Observation of a topological phase transition in numerics}

\begin{figure}
\centering
\includegraphics[width=15cm]{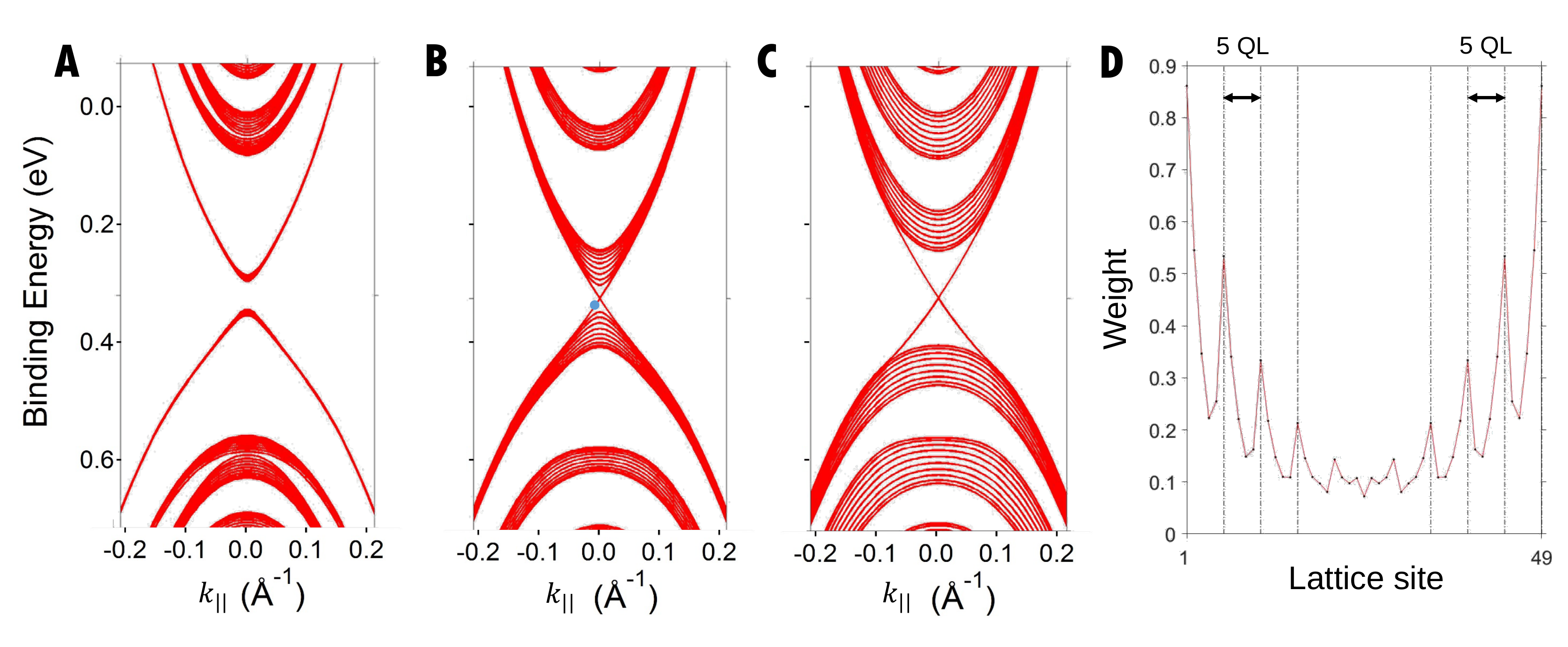}
\caption{\label{FigS1}\textbf{Topological phase transition in numerics.} Dispersion of a tight-binding model taking into account hopping between Se $p$ orbitals in a topological insulator heterostructure. Changing the hopping amplitudes shows (\textbf{A}) a trivial insulator with gapped superlattice band, (\textbf{B}) a phase near the critical point, with a superlattice band at the critical point and (\textbf{C}) a topological insulator with a gapped superlattice band and gapless surface state. (\textbf{D}) Wavefunction of a superlattice state, showing that the state is spatially localized at the interfaces of the heterostructure. The plotted state is marked by the blue circle in (\textbf{B}). Note that in this model we do not \textit{a priori} assume any topological surface or interface states.}
\end{figure}

We perform a tight-binding calculation of the band structure of our topological insulator heterostructure. We consider only hopping between Se $p$ orbitals within the Se planes and between the Se planes of Bi$_2$Se$_3$, with hopping $t$ within the topological layer, $t'$ within the trivial layer and $t''$ at the interface between layers. By tuning the hopping parameters, we observe a band inversion between the two superlattice bands, associated with a topological phase transition between a topological insulator and a trivial insulator, shown in Fig. \ref{FigS1}. We also plot the wavefunction of a specific state in real space, showing that the state is localized only at the heterostructure interfaces.

\section{3. Fine dependence on Indium doping}

\begin{figure}
\centering
\includegraphics[width=14cm]{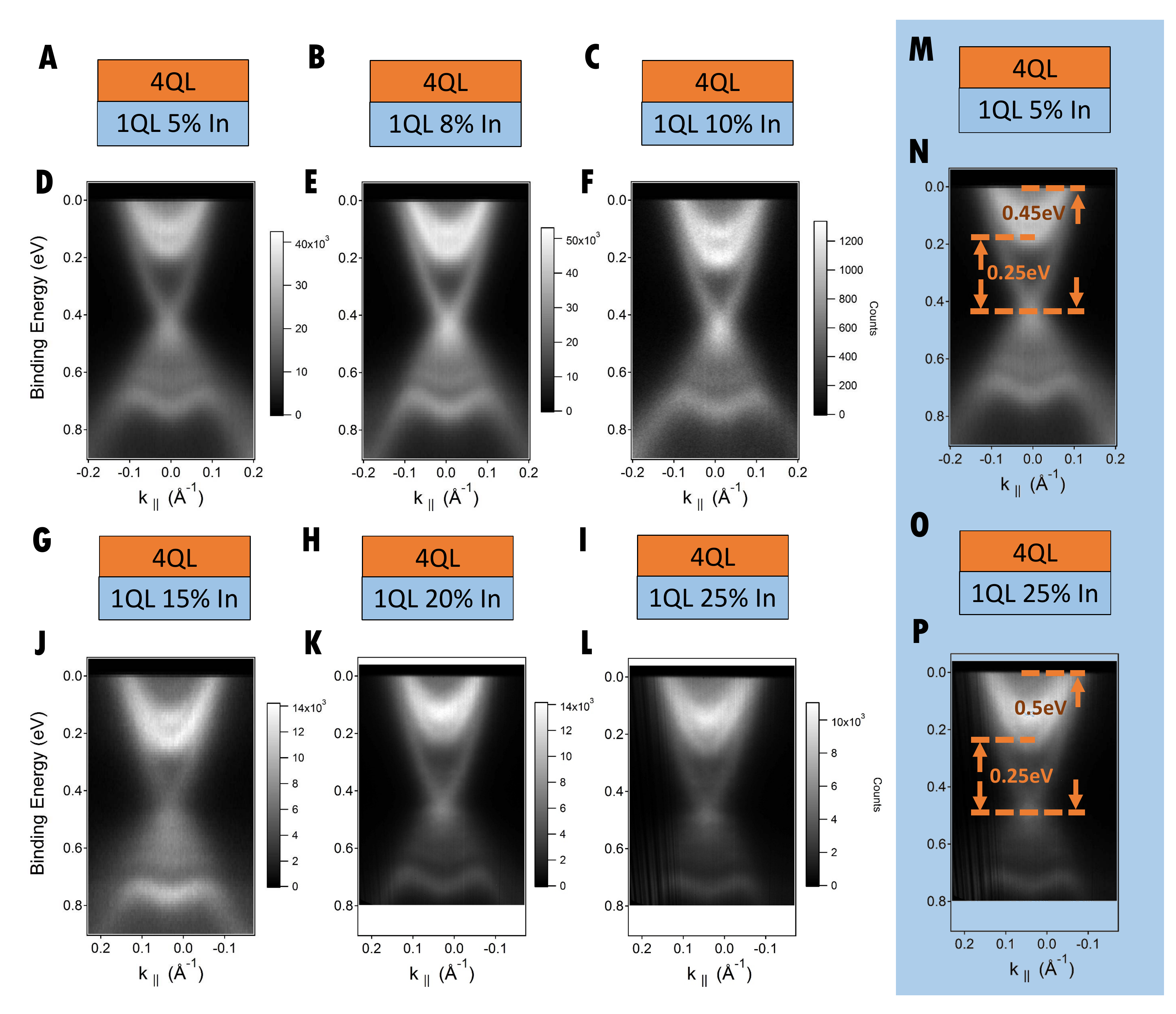}
\caption{\label{FigS2}\textbf{Change in Fermi level with In doping.} Systematic study of heterostructures with compositions shown in (\textbf{A}-\textbf{C}) and (\textbf{G}-\textbf{I}). All samples have 4QL topological layers and 1QL trivial layers, but have varying In doping in the trivial layer. (\textbf{D}-\textbf{F}, \textbf{J}-\textbf{L}) ARPES spectra, showing an $E$-$k$ cut through $\bar{\Gamma}$. We see that the bulk valence and conduction band quantum well states do not change as a function of In doping. In particular, the bulk band gap does not decrease despite a 20\% increase in In doping in the adjacent trivial layer, showing that there is no observable In diffusion to the topmost topological layer. At the same time, the Fermi level rises further into the bulk conduction band. This $n$ doping clearly increases with In doping and we speculate that it is due to the large band gap of the trivial layer under large In doping. Blue panel: we show again the first and last composition in this series, with 5\% In (\textbf{M}) and 25\% In (\textbf{O}). (\textbf{N}, \textbf{P}) The distance between the conduction band minimum and the Dirac point is $\sim 0.25$ eV for both compositions, but the Fermi level moves upward, so the Dirac point is $\sim 0.45$ eV below the Fermi level for 5\% In and $\sim 0.5$ eV below the Fermi level for 25\% In. We see that while the bulk band gap of the topmost topological layer does not begin to close, the In doping in the trivial layer is different in the different samples. This shows that the heterostructure consists of sharp interfaces.}
\end{figure}

We provide an additional, independent check that there is no In diffusion at least near the top of our heterostructure. We present $E$-$k$ cuts through $\bar{\Gamma}$ of 4QL/1QL heterostructures with In doping 5\%, 8\%, 10\%, 15\%, 20\% and 25\% at an incident photon energy of $h\nu  = 18$ eV, shown in Fig.  \ref{FigS2}. We consider the separation between the inner bulk conduction quantum well state and the Dirac point and we observe that this energy difference is $\sim 0.2$ eV independent of the In doping. We make similar observations about the other bulk conduction quantum well state as well as the bulk valence quantum well states. We note that bulk Bi$_2$Se$_3$ is very sensitive to In doping, with a topological phase transition at only $\sim 4\%$ In. However, the trivial In doping in these heterostructures changes by 20\% without any gap closing in the quantum well states of the top layer. This shows that In does not diffuse into the topmost topological layer from the topmost trivial layer. At the same time, we observe that the Fermi level rises with increasing In doping. This rise in the Fermi level is again inconsistent with In diffusion into the \topo\ layer because the bulk conduction band recedes above the Fermi level with increasing $x$ in \triv\ [12]. We suggest that the rise in the Fermi level with In doping may result from the large band gap of the topologically trivial layers. Specifically, the topological and trivial layers have different carrier density, leading to a band-bending effect that may correspond to an effective $n$ doping on the top \topo\ layer. Regardless of the mechanism of $n$ doping, we note that this systematic shift of the Fermi level with In doping confirms that the In concentration indeed increases in the trivial layers. The dependence of the bulk gap and Fermi level on In doping shows that the heterostructure consists of high-quality interfaces.

\section{4. Comparison with a single thin film of B\lowercase{i}$_2$S\lowercase{e}$_3$}

\begin{figure}
\centering
\includegraphics[width=13cm]{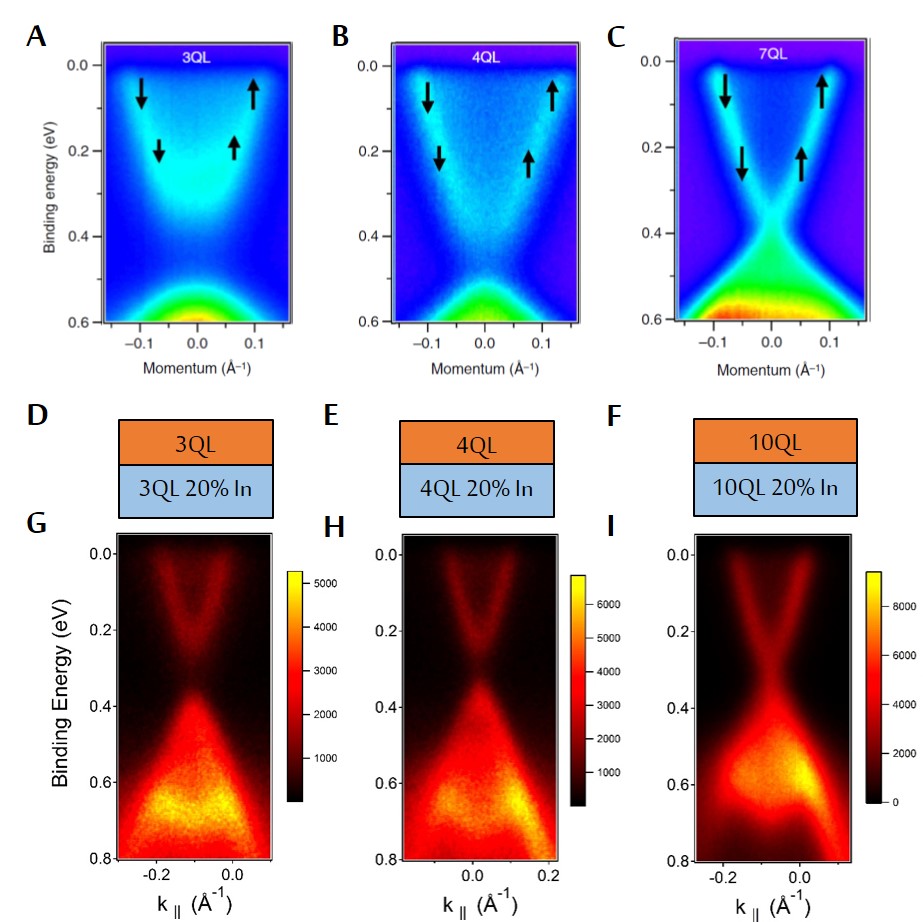}
\caption{\label{FigS3}\textbf{Dimerized-limit heterostructure and a single thin film of Bi$_2$Se$_3$.} (\textbf{A}-\textbf{C}) Gapped topological surface states in a single thin film of Bi$_2$Se$_3$, of thickness 3QL, 4QL and 7QL, respectively, adapted from [14]. (\textbf{D}-\textbf{F}) Unit cells of heterostructures \samG, \samA\ and \samH. (\textbf{G}-\textbf{I}) ARPES spectra, showing an $E$-$k$ cut through $\bar{\Gamma}$. The band gap is $\sim 0.2$ eV in the single 3QL thin film, but $\sim 0.15$ eV in \samG. The band gap is $\sim 0.1$ eV in both the single 4QL thin film and \samA. The surface state is gapless for the single 7QL thin film and \samH. The difference in the band gap at 3QL cannot be attributed to In diffusion because In diffusion should shrink the effective thickness of the topological layer and increase the size of the gap. The smaller gap must therefore be due to superlattice dispersion. Because the gap in the heterostructure is no larger than the gap in the single thin film, it is clear that there is no In diffusion from the topmost trivial layer into the topmost topological layer and the heterostructure consists of sharp interfaces.}
\end{figure}

We present yet another, independent check that the heterostructure is well-defined by considering dimerized-limit heterostructures. We repeat the $E$-$k$ cuts through $\bar{\Gamma}$ of \samG, \samA\ and \samH, discussed in the main text, and we show earlier results on single thin films of Bi$_2$Se$_3$, from [17]. We see that the size of the gap in the topological surface states is $\sim 0.2$ eV for a single 3QL film of Bi$_2$Se$_3$, but $\sim 0.15$ eV for \samG, as shown in Fig. \ref{FigS3}. Further, the gap is $\sim 0.1$ eV for both \samA\ and a single thin film 4QL thick. Lastly, the gap for a single thin film vanishes above 7QL, and we observe a gapless surface state in \samH. If there were In diffusion from the topmost trivial layer to the topmost topological layer, then the effective thickness of the topological layer should decrease, causing the surface state gap to increase. Since the gap in the heterostructure is no larger than the gap in the corresponding single thin film, there is no In diffusion into the topmost topological layer. Incidentally, the $\sim 0.15$ eV gap in \samG\ therefore shows an emergent superlattice band structure away from the dimerized limit, with small but observable hopping across the trivial layer. Our comparison of topological insulator heterostructures with single thin films of Bi$_2$Se$_3$ again shows that the heterostructure consists of high-quality interfaces.

\section{5. Detailed analysis of bulk quantum well states}

\begin{figure}[h!]
\centering
\includegraphics[width=16cm]{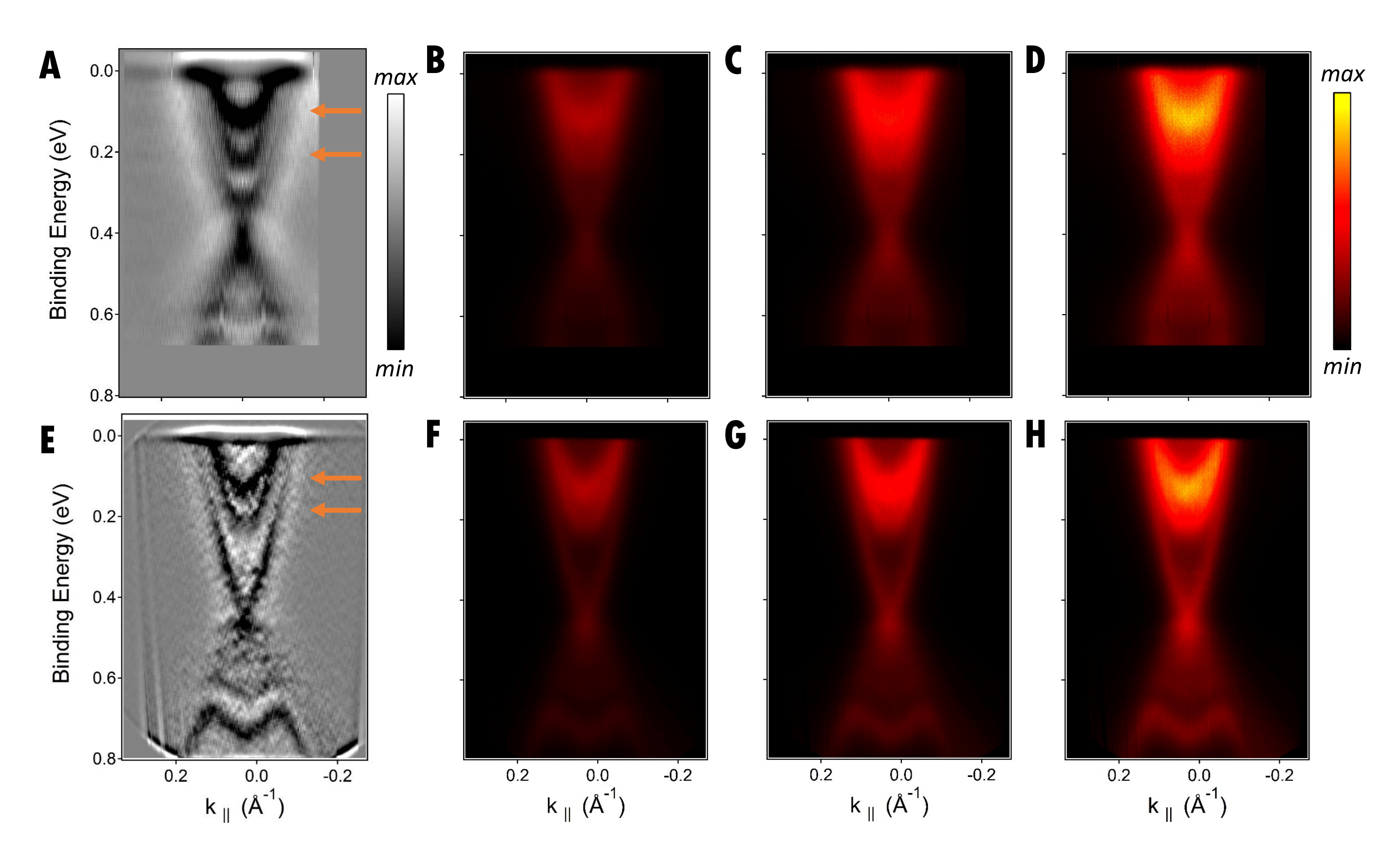}
\caption{\label{FigS5}\textbf{Bulk quantum well states of \samC\ and \samD.} (\textbf{A}) A second-derivative map of the ARPES spectrum of \samC, shown in Fig. 2 G of the main text. We clearly see two quantum well states in the bulk conduction band, marked by the orange arrows. (\textbf{B}-\textbf{D}) The same cut, with different color scales, to make the two quantum well states visible in the raw data. (\textbf{E}) The same as main text Fig. 3 B, repeated here for completeness. Again, we clearly see two quantum well states in the bulk conduction band. (\textbf{F}-\textbf{H}) The same cut, with different color scales, to make the two quantum well states visible in the raw data.}
\end{figure}

To clearly demonstrate the pair of bulk quantum well states in \samC\ and \samD, we present additional analysis of the spectra that were shown in Figs. 2 G, H of the main text. In Fig. \ref{FigS5} A, we show a second-derivative map for \samC, to complement the second-derivative map shown for \samD\ in main text Fig. 3 B. We also show the raw data in several color scales, to make the features more visible. We clearly observe two bulk quantum well states in both spectra, as well as a gapless topological surface state and valence band quantum well states.

\end{document}